# On the uses and abuses of regression models: a call for reform of statistical practice and teaching

John B. Carlin & Margarita Moreno-Betancur

Murdoch Children's Research Institute
&
The University of Melbourne

## *Abstract*

Regression methods dominate the practice of biostatistical analysis, but biostatistical training emphasises the details of regression models and methods ahead of the purposes for which such modelling might be useful. More broadly, statistics is widely understood to provide a body of techniques for "modelling data", underpinned by what we describe as the "true model myth": that the task of the statistician/data analyst is to build a model that closely approximates the true data generating process. By way of our own historical examples and a brief review of mainstream clinical research journals, we describe how this perspective has led to a range of problems in the application of regression methods, including misguided "adjustment" for covariates, misinterpretation of regression coefficients and the widespread fitting of regression models without a clear purpose. We then outline a new approach to the teaching and application of biostatistical methods, which situates them within a framework that first requires clear definition of the substantive research question at hand within one of three categories: descriptive, predictive, or causal. Within this approach, the development and application of (multivariable) regression models, as well as other advanced biostatistical methods, should proceed differently according to the type of question. Regression methods will no doubt remain central to statistical practice as they provide a powerful tool for representing variation in a response or outcome variable as a function of "input" variables, but their conceptualisation and usage should follow from the purpose at hand.

---

Corresponding author: John Carlin, Murdoch Children's Research Institute & University of Melbourne, Flemington Road, Parkville, VIC, Australia 3052. Email: jbcarlin@unimelb.edu.au

## *Introduction*

This article is premised on the fact that while regression methods dominate much of biostatistical practice, many applications of regression analysis in the medical and health research literature lack clarity of purpose and exhibit misunderstanding of key concepts. Although we focus on regression methods, because of their central place in the mindset of applied statisticians, our concerns are relevant to biostatistical methods more broadly and are more fundamental, going to the essence of what statistical thinking and statistical concepts bring to empirical science and how this is taught. We situate our discussion within the specific confines of the discipline of biostatistics, which encompasses the concepts and methods that underlie quantitative research studies in epidemiology and clinical medicine. Indeed, a central theme of the paper is that the teaching of statistical concepts and methods needs to be deeply embedded within their intended area of application, so although we are confident that our main ideas are relevant to applied statistics beyond biostatistics, we will leave it to others to make those connections and extensions. To be more specific, our concerns focus on the biostatistical concepts and methods relevant to data analysis activities that seek to infer new knowledge about a target (human) population of interest. This could be in an area where little is known, with the investigation seeking to inform further research, or in one where much is known, in which case the investigation might aim to confirm prior studies.

Recent developments in epidemiological methodology have highlighted the importance of clearly distinguishing "three tasks of data science", or three types of research question.[1] It can be argued that each research data analysis task has either (i) a descriptive purpose – characterising the distribution of a feature or health outcome in a population,[2] (ii) a predictive purpose – producing a model or algorithm for predicting future values of an outcome given individual characteristics,[3] or (iii) a causal purpose – investigating the extent to which health outcomes in a population would be different if a particular intervention were made.[4] While each specific analysis pursues a question of one of these types, it should be emphasised that most areas of health and medicine advance by examining questions of all three types. Unfortunately, this fundamental taxonomy of research questions has barely penetrated the teaching and practice of biostatistics, especially with respect to regression methods. Indeed, biostatisticians continue to teach, and users of biostatistical methods continue to internalise, the idea that regression models provide an all-purpose toolkit that can be implemented more or less agnostically to the actual purpose. A widespread approach is first to "find the best model for the data" and second to develop an appropriate interpretation of the fitted model.

Others have pointed to some of the deficiencies and hazards of this approach. For example, Westreich and Greenland[5] coined the term "Table 2 fallacy" for the still common practice of presenting a table of estimated regression coefficients from a multivariable model with the implication that these coefficients represent usefully interpretable quantities. Such presentations commonly suggest or imply a causal interpretation, i.e. that changing the value of a variable (while "holding all other variables constant") would lead to a change in the outcome of a magnitude represented by the variable's regression coefficient. As Westreich and Greenland point out, valid estimation of a causal effect requires the delineation of a range of assumptions, both causal and parametric, and there are no reasonable assumptions under which the coefficients of a multivariable regression model simultaneously provide estimates of the causal effects of every variable in the model. If this is understood, it is a short step to ask to what questions, if any, the coefficients of these ubiquitous models provide answers.

Related to the last point is the widespread application of regression methods for addressing vaguely framed questions such as "what are the important risk factors for condition *Y*?" where what is meant by "risk factor" remains ill-defined, often encompassing a combination of potential causes and predictors.[6,7] Such applications imply that it is meaningful to fit a multivariable regression and use data-driven variable selection to reduce the candidate list of risk factors to those that are found to have "independent effects" (another ill-defined term), after which process the remaining risk factors are deemed "important". However, the purpose for which they might be "important" is invariably unclear – with this approach, it would not be as intervention targets nor for outcome prediction.

These practices reflect what we describe as the "true model myth": the notion that the statistical analyst's primary task is to identify a model that best describes the variation in an outcome in terms of a list of "independent variables". Finding the best model is rapidly conflated with the idea that the identified model provides a useful approximation to the actual data generating process – from which empirical conclusions can then be drawn. Textbooks and courses are dominated by these notions, reflecting the traditional pedagogical approach in statistics of presenting techniques and related theory ahead of key questions about their practical application.

The importance of clarity of purpose in the use of statistical models has been emphasised by many others of course, often with reference to Box's famous aphorism "All models are wrong, but some models are useful…"[8] However, there seem to have been few if any attempts to provide an in-depth examination of how the "usefulness" of a model might be defined and how the teaching and practice of statistical analysis might change accordingly. Hernán et al[1] briefly describe a similar agenda in the broader context of "data science" but do not explicitly address the central role of and difficulties raised by regression methods, nor the potentially key role of statisticians in tackling these issues.

We begin this paper by briefly introducing some standard notation and then describe the problems with current practice in greater detail by way of three examples from the first author's own practice. After analysing the heart of the problem and its continuing prevalence, via a brief review of mainstream clinical research journals, we present a proposal for the reform of teaching and practice. This takes the form of an outline of how biostatistical methods could be taught from a "purpose-driven" perspective, using the taxonomy of the three types of research question, and introducing the technical aspects of regression models and other methods when relevant to the purpose.

## *Notation and core concepts of regression models*

For later convenience and clarity of terminology we begin with an overview of basic concepts and notation related to regression analysis, noting that at this stage we remain agnostic to its purpose and simply set out standard mathematical definitions that we can refer back to. For a continuously varying outcome or response variable $Y$ and a single covariate $X$ in some well-defined population, we use the following standard notation for the simple linear regression model:

$$Y = \beta_0 + \beta_1 X + \epsilon \,,$$

where $\beta_0$ represents the expected (average) value of $Y$ when $X = 0$, $\beta_1$ represents the difference in the expected value of $Y$ between individuals in the population for whom the values of $X$ differ by one unit, and $\epsilon$ represents a zero-mean "error" or deviation of $Y$ from the expectation $\beta_0 + \beta_1 X$. The covariate $X$ may be a dichotomous indicator, in which case $\beta_1$ encodes the difference in the mean value of $Y$ between two subgroups, or it may have multiple values, in which case the model incorporates an assumption of linear increase in the mean value across the values of $X$. The random error $\epsilon$ is commonly assumed to follow a normal distribution with constant variance $\sigma^2$, with errors for each individual independent of each other.

The essence of the model is better represented by separating the expression for the expected value (conditional on $X$), from the assumptions about the error term or probability distribution:

$$\mu(X) := E(Y|X) = \beta_0 + \beta_1 X; \;\; Y|X \sim N(\mu(X), \sigma^2). \tag{1}$$

In most contexts, however, there is not just a single $X$ that may predict the value of $Y$, and in addition, not all outcomes of interest lend themselves naturally to modelling their expected value in this way, especially when they are not continuous. Mathematically, it is appealing to address the former of these issues by extending (1) to the *multivariable* (more traditionally termed *multiple*) regression model, with the following expression for the expected value:

$$\mu(\boldsymbol{X}) := E(Y|X_1, X_2, \dots, X_p) = \beta_0 + \beta_1 X_1 + \beta_2 X_2 + [\dots] + \beta_p X_p \tag{2}$$

(where the bold $\boldsymbol{X}$ signifies a vector or list of covariates $X_1, X_2, \dots, X_p$). A fundamental feature is the *linear* nature of the model, i.e. linear in the coefficients, which lends itself to convenient matrix representations and a related range of appealing mathematical properties. Importantly, the mean specification in the canonical multiple regression model (2) may include complicated (non-linear) functions of the original measured variables, including interaction terms, i.e. products of original variables.

The second of the two issues flagged above is readily addressed by introducing the concept of the *link function*, allowing the linear predictor on the right-hand side of (2) to be specified as a model for a non-linear ("link") function $g$ of the expected value:

$$g(\mu(\boldsymbol{X})) = g(E(Y|\boldsymbol{X})) = \beta_0 + \beta_1 X_1 + \beta_2 X_2 + [\dots] + \beta_p X_p. \tag{3}$$

This *generalised linear model* (GLM) formulation allows the ideas of regression analysis to be extended to non-continuous outcomes and non-normal error distributions. Classic examples include logistic regression for a binary outcome $Y$, in which the link function is $g(\pi) = \text{logit}(\pi) = \log(\pi/(1-\pi))$, where $\pi = \Pr(Y = 1)$. This extension of the basic regression concept is facilitated by the separation in (1) between the "fixed" part of the model (the expression for the expected value or the link-function-transformed expected value) and the "random" part, which specifies the nature of the variation around the fixed part.

## *Common uses and abuses of regression in empirical research*

Regression analysis is ubiquitous as the pre-eminent tool of applied biostatistics. Data analysts develop a reflexive instinct that their role in many settings is to identify an appropriate, "well-fitting" regression model for the data. Unfortunately, the commonly observed lack of specificity with respect to the actual aim of many analyses leads to a range of pitfalls and difficulties.

We illustrate these issues by describing three examples of published analyses on which the first author collaborated early in his career. Following initial description of the examples in this section, the next section further examines the problematic issues that they raise.

**Example 1**: Acute pyelonephritis and kidney enlargement in young children[9]

Young children who acquire a urinary tract infection may also develop an infection of the kidney known as pyelonephritis. Affected kidneys become enlarged during these infections, which makes it difficult to use ultrasonic measurements of kidney size as a reliable baseline for future assessment of growth. The research described here sought to estimate how much affected kidneys were enlarged, compared to normal kidneys. Clinical researchers systematically measured kidney length using ultrasound scans in a consecutive series of 180 children diagnosed with urinary tract infections and referred for imaging between February 1990 and June 1992 at a major paediatric hospital in Melbourne. Of these children, 77 had also developed pyelonephritis (according to a nuclear medicine scan called a scintigram) in one or both kidneys. The children ranged in age from newborn to just under 5 years, and 58% were (biologically) female. Given this variation in age (especially) and sex, it is not immediately clear whether it is meaningful to characterise the extent of kidney enlargement in those who developed pyelonephritis compared to those who did not as a single summary number: perhaps an adequate description would require separate analysis by categories of age and sex. However, an initial examination of the data broken down by age (in years) and sex indicated that the difference in averages between pyelonephritic (infected) and normal (uninfected) kidneys was roughly constant across these categories. This in turn suggested to the biostatistician consulted to assist with analysis (JBC) that a regression model would provide an effective tool for estimating the difference in means between infected and uninfected kidneys at any age, with "adjustment for age" accomplished by including a smooth function of age as a covariate in the model. The original analysis also "adjusted for sex" and allowed for correlation between kidney lengths in the same child (170 children had measurements on both kidneys while for 10 a single kidney was measured), but for simplicity of exposition we ignore these details.

Thus, in essence, the analysis comprised fitting a linear regression with mean specification:

$$\mu(X_1, X_2) = E(Y|X_1, X_2) = \beta_0 + \beta_1 X_1 + \beta_2 X_2 + \beta_3 X_2^2 + \beta_4 X_2^3 \quad (4)$$

where $Y$ = kidney length (mm), $X_1$ = 0/1 indicator for infection status and $X_2$ = age (months). This model encodes two key assumptions: that the difference in average kidney length between infected and uninfected kidneys is a constant ($\beta_1$) for all values of $X_2$, while for both values of $X_1$ the average length increases by a cubic function of age, determined by the coefficients $\beta_2$, $\beta_3$, $\beta_4$.

The main results can be seen in the form of two smooth curves displaying the fitted values for the infected and uninfected groups (those with and without a "scintigraphic defect") from this regression model overlayed on a scatter plot of the raw data (Figure 1). From the raw data, it's difficult to see a clear difference between the infected and uninfected kidneys (although see Supplementary Material for a clearer representation using colour and a better choice of plotting symbols), but the fitted curves portray, as per the model's underlying assumption, a constant mean difference between the two groups, estimated to be 4.1mm. The analysis has a certain prima facie appeal even though the rationale for "age adjustment" was unclear in the original article; see next section for further discussion.

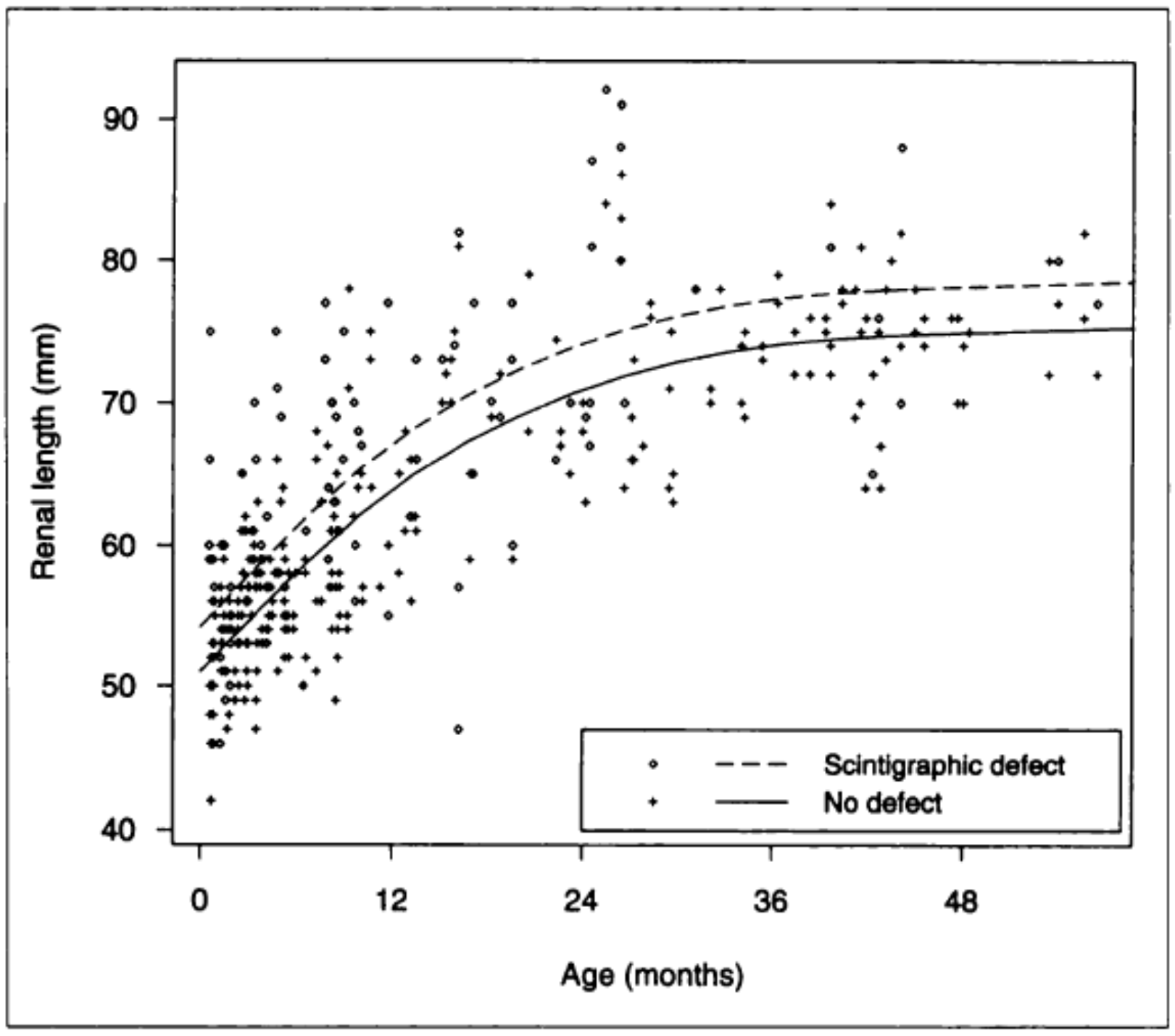

**Fig. 1.—Scatter plot of renal length measured on sonograms versus age for kidneys with and without defects shown on scintigrams. Curved lines represent cubic model used in analysis of covariance calculations. Because curves are parallel, there is a similar absolute increase in renal length at all ages.**

**Figure 1**: Key figure from paper on kidney length and pyelonephritis in young children (Example 1).[9] Note that for this analysis "Scintigraphic defect" was taken to be synonymous with kidney infection (despite occasionally reflecting non-infectious scarring). Also note that the final statement in the figure caption was confusing because it obscures the fact that the curves are only parallel because they were derived from a model in which they were *assumed* to be parallel (see text).

**Example 2**: Predicting the success of gas enema treatment for intussusception in children[10]

Intussusception is an acute bowel constriction that occurs infrequently in very young children. It is painful and can be dangerous because it blocks the intestines, so surgeons are called upon to intervene and relieve the blockage. The standard treatment (at the time of the research described here) was to use a "gas enema", a simple procedure that injects air into the baby's rectum. The procedure is usually successful, but not always, so the clinical investigators of this study were aiming to understand the extent to which a successful outcome could be predicted using characteristics of the child or their clinical presentation. Data were collected prospectively on 282 consecutive cases of intussusception that were treated with gas enema, after presenting to a major tertiary paediatric hospital between January 1987 and July 1991, and included an indicator for the outcome (success or failure) and a set of clinical covariates that were candidates for predicting the outcome.

Given the binary outcome measure, a natural statistical approach to prediction is to build a logistic regression model. In this case, the biostatistician (JBC) used a backwards selection procedure to reduce the number of candidate variables in the regression specification to those that seemed to be important, according to conventional (though increasingly deprecated[11,12]) criteria based on "statistical significance"; in this case, the usual threshold of $p<0.05$ was used. They also followed standard procedures in sticking to simple specifications that ignore potential interaction terms between covariates.

The main results were displayed in a table (Figure 2) that presented estimates with confidence intervals and p-values for the coefficients that remained "statistically significant" in the multivariable model. Although the predictive purpose of this paper was clear, there remain several issues with the way regression analysis was used to address this; see next section.

**TABLE 4: Results of Logistic Regression Analysis with Successful Gas Enema as Outcome for Children with Intussusception**

| Predictor Variable[a] | *p* Value[b] | Odds Ratio[c] | 95% Confidence Interval |
|---|---|---|---|
| Dehydration level | <.001 | | |
| 1–4% | | 0.32 | (0.13, 0.80) |
| 5% | | 0.13 | (0.05, 0.33) |
| 6–10% | | 0.10 | (0.02, 0.42) |
| Duration of symptoms >12 hr | .03 | 0.42 | (0.02, 0.90) |
| Small-bowel obstruction | .005 | | |
| 1–2 fluid levels | | 0.78 | (0.32, 1.90) |
| >3 fluid levels | | 0.24 | (0.10, 0.57) |
| Palpable mass present | .03 | 2.43 | (1.07, 5.50) |

Note.—Baseline odds of successful gas enema for well-hydrated patients who had signs and symptoms for less than 12 hr, no obstruction, and no palpable mass were 10.1.

[a]Variables in Table 1 that are omitted from this table showed no significant contribution to the multivariate model. No significant interactions were found between the independent variables.

[b]Likelihood ratio test for variable when entered last.

[c]Odds ratio comparing given level of each variable with baseline: for example, at dehydration level 1, odds of success are .32 times the odds for normal hydration, assuming all other variables remain constant.

**Figure 2**: Reproduction of key table presenting the results of a multivariable logistic regression analysis for the prediction of successful gas enema in children with intussusception (Example 2).[10] See text for discussion.
[Reproduced with permission of the American Roentgen Ray Society from "Gas enema for the reduction of intussusception: relationship between clinical signs and symptoms and outcome", Katz M et al, American Journal of Roentgenology, 160, 363-366. Copyright© 1993 American Roentgen Ray Society.]

**Example 3**: Associations between multiple risk factors and risk of childhood asthma[13]

In this example, the stated aim was to estimate the "strength of association" of numerous potential "risk factors" with the risk of childhood asthma. The data derived from a cohort study of all children born in 1961 who were attending school in Tasmania in 1968 (aged 7 years), with this paper reporting a cross-sectional analysis of data collected from the parents at the time of recruitment ($n$ = 8585). Questionnaires were used to determine both the primary outcome of interest (history of asthma in the child) and the risk factors, including child's sex, other atopic conditions (such as hay fever and eczema), family history of allergic disease and parental smoking.

As with the previous example, a binary outcome triggered the use of logistic regression. The paper states that "[f]ollowing Hosmer and Lemeshow, a parsimonious model with no interactions was determined by forward selection", with reference to the first edition of a well-known textbook.[14] Because of the large sample size, some risk factors with quite small estimated regression coefficients survived the selection process (adjusted to a lower-than-conventional significance threshold of p<0.01). Results were presented in the table shown in Figure 3. The main-text presentation of results also included the reporting of so-called "crude associations", with some discussion of the way in which the differences between crude and "adjusted" estimates reflected likely "confounding" effects. It was also reported that "there were no differences between the ORs in males and females", and in fact all 55 potential two-way interactions were reported to be non-significant at the 0.01 level. The opening summary of the paper's Discussion stated that "These atopic conditions [history of various other allergic conditions] were found to be independent risk factors, in that an increased risk of asthma was associated with each factor even though the increased risks associated with all other factors had been taken into consideration by the statistical model." It is unclear what this means, as mentioned in the Introduction and further discussed below.

**Table 2. Odds ratios and 99% confidence intervals for child's asthma after adjustment for all other factors in the model**

| Risk factor | *n*=7368 |
|---|---|
| Maleness | 1.56 (1.30–1.86) |
| Hay fever | 3.86 (3.12–4.78) |
| Eczema | 2.04 (1.63–2.55) |
| Hives | 1.34 (1.09–1.65) |
| Allergy to foods or medicines | 1.70 (1.26–2.30) |
| Maternal asthma | 2.63 (2.08–3.31) |
| Paternal asthma | 2.52 (1.99–3.19) |
| Maternal smoking | 1.26 (1.05–1.51) |

**Figure 3**: Reproduction of key table presenting the results of a multivariable logistic regression analysis for the risk that a child has asthma (Example 3).[13] See text for discussion.

## *The essence and extent of the problem*

As discussed in the Introduction, it has been convincingly argued that the purpose of a data analysis task in an epidemiological research context can be classified into one of three distinct types of inquiry or research question: descriptive, predictive or causal.[1] Beyond these three categories, it is difficult to conceive of other research purposes for which an analysis might be intended. If the type of a proposed research question is not immediately clear, this signals that more work is needed to bring it into sharper focus.

How do our three examples fit into the trichotomy? In the first example, the purpose was fundamentally descriptive, aiming to characterise the difference in average kidney length between children whose kidney(s) had and had not been infected in this population. Unfortunately, the rationale for and underlying assumptions of the regression analysis used were not clearly articulated in the paper. In particular, the key simplifying assumption underlying the model, of a constant shift in the distribution between infected and uninfected across the age range, is unlikely to hold exactly in the population. However, by introducing this assumption, with a risk of bias associated with the extent to which the assumption is incorrect, the model provides enhanced precision in the estimated mean difference of interest, by accounting for extraneous variation due to age (see Supplementary Material).

The second example posed a research question that was essentially predictive: how can clinical factors be used to predict successful gas enema for intussusception in young children? The analysis presented a fitted multivariable logistic regression model identified by a variable selection approach that would no longer be recommended in this predictive context. Importantly, no formal assessment of predictive ability or external validity was conducted, with these aspects only mentioned briefly in the paper's Discussion. Furthermore, the presentation of estimated regression coefficients in the table shown in Figure 2 (and its footnote a) erroneously implied that the analysis demonstrates that the covariates included (and the way they were included, i.e. with no interactions) are the only useful predictors, among the available covariates. It also implied that the estimated odds ratios displayed in the table have a meaningful interpretation (see footnote c), which is surely not the case: they are simply coefficients that might be used within a prediction algorithm (as was in fact illustrated briefly in the Discussion).

In the final example, the authors' purpose was vague, but when examining the motivation and conclusions of the study, it appeared to be essentially causal[15]: the aim was to identify "risk factors" that might be useful for informing intervention strategies to reduce the risk of asthma. In this context, the reported regression analysis is difficult to interpret, without incurring the "Table 2 fallacy".[5] Critically, a clear causal intent requires attention to the precise definition of putative causal exposures, for example considering the extent to which they might be modifiable and if so how, because interventions to modify exposures are what causal questions seek to inform.[16]

A common feature of the uses of regression in these examples is that the analyst's purpose was not fully articulated, with each analysis based on an implicit assumption that statistical methods can be used to construct the "best model" for the outcome given the available covariates, from which relevant conclusions can then be drawn. In effect, the model is understood to provide a representation of the true data generating process, revealing how the independent variables combine to "produce" the outcome. This approach might work if it were empirically the case that processes in the real world of population health and medicine, such as those underlying these examples, obeyed natural laws of the form that regression models represent, *and* that researchers were able to measure all relevant variables, but this is surely far from the reality.

The examples described so far reflect some of the first author's applied practice in the 1990s, but we observe that similar practices continue to be widespread today. Over the past several decades the technical complexity of statistical analysis presented in medical journals has increased markedly, with much greater use of multivariable regression analysis. For example, one review noted a ten-fold increase (from 5% to 51%) between 1978-1979 and 2004-2005 in the proportion of articles using multivariable regression in the *New England Journal of Medicine.*[17] Another reported a doubling (36% to 78%) in the use of "regression models" in four psychiatric research journals between 1996 and 2018.[18] There appear to be several reasons for the popularity of multivariable regression, including the growth in availability of user-friendly statistical software, in tandem with a widespread belief that multivariable regression is an omnibus tool for addressing many problems.

A systematic review of contemporary usage of multivariable regression is beyond the scope of this paper but we provide a brief targeted review covering a single month of publication (June 2022) in three leading journals of clinical research: *Pediatrics*, *Neurology* and *BMJ Open*. These journals were selected from a listing of the top 20 "most influential medical journals"[19], to represent journals that carry a high proportion of observational clinical studies while leaving aside those (generally the most highly ranked general medical journals e.g. *NEJM*, *Lancet*, *JAMA*) that carry a predominance of large randomised trials and others that publish a greater proportion of laboratory studies. For articles published in the (arbitrarily) selected month, we identified the proportion that reported any form of regression analysis (while briefly characterising the remaining articles) and we classified the identified articles according to (i) whether regression analysis was used for a clear purpose (descriptive, predictive, or causal), and (ii) whether key misuses of regression methods were apparent.

Overall, we examined 57 papers (18-20 per journal), in 36 (63%) of which regression methods were used. Among these papers, 25 (69%, or 44% of all papers) exhibited a type of misuse of regression along the lines that we have identified above (see Supplementary Material for details). Although the mix of types of question was quite different across the three journals, the proportion of papers in which misuse could be identified was similar. The most commonly observed problem was the fitting of multivariable regression models without full consideration of the precise aims of the study, in a manner that exemplifies the "true model myth". Specifically, we found 10 instances of multiple regression applied to ill-posed questions along the lines of "can we identify the [most important] risk factors for [condition Y]?". Furthermore, even when a clear research question was identified, we observed frequent misuse of regression, such as inadequately justified "adjustment for covariates" and erroneous interpretation of estimated coefficients.

The problems identified have clear roots in the way that regression methods are taught. In particular, the multivariable model is invariably introduced as a "natural" extension of simple univariate regression, with the implication that it has widespread inherent applicability and usefulness, while key details of how and why it might be used remain vague. For example, in the introductory chapter of a popular text on logistic regression the authors state that "the goal of an analysis using this model is the same as that of any other regression model used in statistics, that is, to find the best fitting and most parsimonious, clinically interpretable model to describe the relationship between an outcome variable and a set of independent (predictor or explanatory) variables."[14] Chapter 2 is entitled "The Multiple Logistic Regression Model", while in the second sentence of Chapter 3 ("Interpretation of the Fitted Logistic Regression Model") we find that "After fitting a model the emphasis shifts from the computation and assessment of significance of the estimated coefficients to the interpretation of their values." The logic here seems entirely backwards, because unless the model is believed to provide a full and faithful representation of the true data generating process, then its coefficients may be completely uninterpretable. The meaning of model coefficients and indeed the entire model should be clear at the point of model specification, *before* any fitting and estimation takes place.

Against this background, the next section maps out a new approach to teaching biostatistical methods whereby regression and other methods would be taught wholly within the context(s) in which they may be useful, i.e. within each of the three purposes or types of research question.

## *A proposed reform program: teaching regression in context*

We propose that regression analysis should not be taught in standalone courses focused on methods and techniques, but rather that the concepts and methods of regression should be introduced as needed within a biostatistics course program that adopts the framework of the "three types of research question". This would imply a substantially reduced emphasis on discussing regression models as mathematical entities outside of the context defined by the purpose for which they are to be used. Many, though not all, of the concepts and technical details that are covered in the conventional teaching of regression methods would still be important to learn, but they would be presented in a different way, as we outline below.

Firstly, we revisit the notion of the "three tasks" or three types of research question, because a shared understanding of this categorisation is essential to everything that follows. When teaching with this approach we have a specific session on identifying types of research questions, in which we find it useful to focus discussion by asking students (and, by extension, collaborators, when applying the ideas in practice) to reflect on the way they expect their findings to be used by readers or "consumers" of the research[15]. For example, perhaps the work will be used to inform resource prioritisation, based on the scale of a health issue or the extent of discrepancies in outcomes between disadvantaged and wealthy (descriptive purpose)? Or might it assist clinicians to make decisions about which patients to monitor closely or to support in additional ways (predictive)? Or will the work feed into decisions or at least further research about treatment practices or management policies (causal)?

The distinctions are not always crystal clear; for instance, we generally regard a question as descriptive if it seeks to provide a broad characterisation of populations or subpopulations (in the latter case, perhaps with the aim of describing the difference between subpopulations, bordering on a prediction question), while reserving the predictive category for problems that seek to provide accurate individual-specific outcome predictions (such as disease prognosis) usually based on multiple measured predictor variables. For the purposes of this paper, we leave aside other data analytic activities such as preliminary data cleaning and pre-analysis checks, and "exploratory" analysis undertaken without a specific question in mind.

Second, we introduce the notion of a "roadmap to analysis", by which we mean a structured set of steps that the analyst should take (in partnership with substance-matter collaborators) in order to provide clarity about the purpose of analysis, link that purpose to an analysis plan, and facilitate the documentation of all limitations (potential biases) that remain once results have been obtained, in order to inform interpretation. The key components of a structured approach or roadmap (with more detailed summaries available elsewhere, e.g.[20]) are:

(i) Explicit definition of the question in terms of a target parameter(s) of interest (problem specification or definition of estimand(s)),
(ii) Consideration of the sampling and measurement issues that determine whether the estimand is in principle estimable from the study at hand ("identifiability"), and
(iii) Articulation of an analysis plan for estimation of the target parameter (within which there may be a role for fitting regression models and related methods).

The sequence of steps is critically important: the problems we have identified above arise from embarking on modelling and estimation before fully addressing the first two steps. In particular, the first step has been generally missed in traditional teaching and practice, supplanted by an assumption that the "parameter of interest" is a term in a statistical model. This omission has been brought to light by developments in the causal inference literature, which clarified the need for a definition of causal effect that does not depend on a pre-assumed parametric model. The need for a relevant estimand definition that precedes a model specification extends to the other question types. In the space available here, we can only sketch out the key ideas of the first two steps, while focusing on the role of regression modelling within these roadmaps, with the primary aim to reset the conventional thinking of statisticians and users of statistics about when and why such models enter the task of research data analysis.

## Descriptive purpose

Examples of studies with descriptive aims include those that examine the prevalence or incidence of diseases or health conditions in a population or across subgroups of a population; prominent recent examples of this type of study are those that have sought to understand the extent and burden of the COVID-19 pandemic.[21,22] The analysis in Example 1 can also be seen to have pursued a descriptive question, of a rather different kind. Importantly, we distinguish descriptive *research questions*, our focus here, from the descriptive analysis of study participants that is invariably provided in study reports (e.g. "Table 1" of the typical paper, providing summary statistics such as means and standard deviations of continuous variables along with percentage breakdowns into key categories of interest). In the latter context, the purpose is literally to describe the study participants, with no intention to draw inferences about a broader population. In contrast, in studies with descriptive scientific aims, the principles of statistical inference are relevant.

### *The roadmap for descriptive questions*

In many ways, descriptive research questions are the easiest, so they provide a good starting point for learning about the roadmap approach outlined above as well as providing a setting in which some of the technical concepts of estimation and regression models can be introduced. Pursuing the example of COVID prevalence for a moment, a descriptive research question (in 2021, say) might have been to determine the proportion of the population of Australia (the target population) that had experienced a COVID infection. In addition to defining the target population, definition of the target parameter also requires a definition of the outcome, which might be defined as the presence of a specific antibody response in the blood. At the second step, we would need to consider how data were obtained to address the question, with consideration given to the fact that a random sample of the target population was surely impossible (because of the limitations of feasible sampling schemes and the fact that some sampled individuals decline to participate or do not produce a suitable blood sample), and nor was it possible to conduct an error-free test for antibody presence in every sampled individual.

For a hypothetical simpler question in which random sampling and perfect outcome measurement were possible, the target parameter would be clearly identifiable, i.e. a summary statistic in a very large sample would provide an unbiased estimate (and thereby answer the research question). This in turn would lead to a very simple analysis plan based on the same summary statistic in the actual finite sample, accompanied by standard inferential statistics to qualify imprecision due to sampling variability. In real-world studies, identifiability of the actual target parameter is invariably questionable, so important tasks for the statistician (albeit largely beyond the scope of this paper) are to develop analysis plans that address issues of sampling and measurement bias as much as possible, and beyond that to document remaining limitations and provide a framework for suitably cautious interpretation of analytic results.

***Introducing regression models in the context of descriptive inference***

How do regression methods enter the analytic toolkit for answering descriptive research questions? As a starting point for later elaboration, we suggest it is instructive to introduce the general concept of a regression model by way of the "null model", which is the version of (1) (or its generalised version with a non-identity link function as in (3)) in which $X$ is dropped (or, equivalently, fixed at 1 for the entire population). In that case, only one parameter is defined: $\beta_0$, representing the mean (or $g()$-transformed mean) in the population. In the idealised "perfect sampling" world, this mean value is the target estimand for a descriptive question, and we can use this setting to provide the connection between simple random-sample-based estimation of mean values and regression estimation technology.

If the descriptive question instead seeks to describe the difference in mean (or risk, or prevalence, for a binary outcome) across subpopulations or subgroups, say in the simplest case of a binary subgroup indicator variable $X$, coded 0/1, then the regression in (1) can be used as it stands, with the coefficient $\beta_1$ representing that difference, while $\beta_0$ represents the mean in the subgroup $X = 0$. Inference for each of these parameters may be obtained by appropriate variance calculations, which could be examined mathematically in advanced classes, and can be performed in practice by fitting this regression in statistical packages (with the inference for $\beta_1$ equivalent to the classical $t$-test with an estimation focus). Alternatively, if the aim is to describe each of the subgroup means rather than their difference, as might be more natural in descriptive studies, these can still be estimated using the regression just described using estimates of $\beta_0$ and $\beta_0 + \beta_1$, augmented with a method for obtaining the variance of the estimate of the mean in subgroup $X = 1$, i.e. $\beta_0 + \beta_1$. An alternative approach is to reparametrise the regression model by dropping the intercept (i.e. the coefficient $\beta_0$) and including indicators for both subgroups, with the corresponding coefficients then representing the subgroup means. A useful teaching point is that the regression approach enables inferences for the subgroup means using a pooled estimate of the variance within groups, rather than relying on separate estimates of variance from each group.

A natural extension is to the case where $X$ takes multiple values. If these represent $k$ unordered categories (a "nominal" variable), then description of the difference in mean for each of the $k-1$ subgroups relative to a reference group may be achieved in a regression framework by including $k-1$ indicator variables in the regression specification, with each of the resulting coefficients encoding those differences. Alternatively, as for the two-group case, the intercept may be dropped, and a coefficient included for all indicators, representing the subgroup means.

By this point in the curriculum, students may be introduced to commonly used software tools for fitting regression models, with applications emphasising that much of the standard output of these tools (such as estimates of model coefficients that do not correspond to target parameters of interest, and breakdowns of sums of squares, including R-squared) is of limited value for practical purposes.

***Regression lines and curve-fitting for descriptive questions***

Another type of descriptive aim is to characterise the variation in the mean of an outcome with a continuous covariate. This provides a point of contact with the traditional introduction to linear regression that focuses on continuous $X$ and continuous $Y$. Univariable linear regression provides a method for describing the joint variation of $X$ and $Y$, focussing on the (asymmetrical) question of how the expected value of $Y$ changes according to the value of $X$. Under the assumption of a linear relationship, the regression equation provides a summary of the average rate of change in the mean of $Y$ with each unit of difference in $X$, smoothing over the variability of individual values to allow the essential strength of (linear) relationship to be estimated.

Of course, there are many examples in which summarising the relationship in the form of the best-fitting straight line is too simplistic. These provide an opportunity for students to learn about "curve fitting", which may be accomplished by using polynomial functions, as in (4), or by more modern alternatives, from parametric fractional polynomials to semi-parametric splines etc.[23,24] Several techniques can thus be discussed within the descriptive context, although it may emerge that their true value becomes debateable as they become more complex – as the "summary measure" describing the data may become scarcely less complex than the raw data themselves. An important related issue with continuous outcomes is the choice of summary measure, in relation to the distribution of the outcome. If the distribution is skewed, should the target parameter of interest still be the mean? Perhaps so, in the analysis of healthcare costs[25], for example, but perhaps not in other settings, where for example the mean in a log scale or the median might be more appropriate. Such considerations need to be guided by the substantive context as well as appropriate data visualisation.

***Introducing multivariable regression***

We have seen that the kidney lengths study (Example 1) used a multivariable regression analysis. Why was this done, and could this be a valuable teaching example? First, as in many clinical studies, the definition of the relevant population was implicit rather than explicit, so there was a fundamental lack of clarity in the actual research question. Leaving that to one side for now (see next section), as well as the issue of potential measurement bias, and proceeding with an analysis based essentially on the "random sample from the population" idealisation, there remain some things that can be learned about statistical concepts and techniques from this example.

To begin with, it should be clear from basic biology, as well as a cursory look at the data, that the outcome depends strongly on age, so it is natural to ask whether the research question is well-posed without considering age. If the infection-related kidney enlargement changed with age, then its extent should be described as a function of age. If, as appeared to be the case, there was a near-constant difference across the age range, then an analysis that *smooths out* the additional outcome variation due to age, for example by obtaining the mean difference within age categories and averaging these over the categories, can be shown to produce an estimate with lower variance than the crude mean difference ignoring age (see Supplementary Material).

The multivariable linear regression model shown in (4) can then be introduced as a convenient and effective tool for smoothing, producing an estimate of the difference between the groups that reduces the variance associated with age even further. In the teaching context, one should emphasise that this tool is essentially a convenient way of containing the variability of the data in order to estimate more precisely what is *assumed to be* a constant mean difference (represented by the parameter $\beta_1$ in (4)) across age, which itself is associated with rapid change in the outcome mean. The assumption of a constant difference across the age range can be checked to some extent (limited by sample size) from the data. The key "outputs" for teaching purposes are the estimate and standard error (or confidence interval) for the coefficient $\beta_1$ in model (4) and in the univariable model (1). An important aside is that this rationale of improved precision via multivariable regression only applies to continuous outcomes, for which variation is independent of the mean.

It should be noted that poor modelling of the dependence of the outcome on age may lead to a biased estimate of the group difference of interest. This would be an example of trading increased bias for a reduction in variance. In fact, for teaching purposes it might be interesting to explore this by fitting a simpler model in which the age relationship is represented by a simple linear trend. In this regard, the example also provides an opportunity to observe the value of nonlinear functions in a regression expression, with the interesting twist that the authors used a cubic polynomial to represent the age dependence. This was done not because the cubic term in the regression specification was "statistically significant" (it wasn't), but to avoid the unrealistic shape of the simplest alternative to a straight line, the quadratic.

***Regression adjustment in descriptive inference***

Although the kidney study example provides a nice opportunity for introducing the basic concepts of multivariable regression, an arguably more important use case is for addressing the ubiquitous threat of sampling bias. The simple analyses described thus far only make sense if the analytic sample is (equivalent to) a random sample of the target population. This is rarely the case, of course, so where possible analysis planning should seek to develop an approach that "adjusts" for differences between the study sample and the population.[2] In the kidney study, no population reference data were available, so there was little scope for "adjustment". In larger epidemiological studies, however, data are often available on covariates that characterise the sample (in terms of age, sex, geography and other sociodemographics), and if the corresponding characteristics of the population are also available then there is scope for reweighting the sample mean values to the population distribution of covariates – a form of "adjustment". This can be done using classical sample survey methods, as discussed in the literature on standardisation of rates in epidemiology[26,27] and seen for example in a recent COVID-19 study[22]. However, standardisation or reweighting to a population can also be facilitated by multivariable regression models[28,29], although most discussions of this topic focus on analysis for causal questions[27,30], and the details may be too complex for this early stage of our proposed curriculum. Importantly, though, even at the introductory level, a full discussion of descriptive research questions should highlight that the role of multivariable regression analysis in this arena is often exaggerated, with frequent examples of "regression adjustment" that is not clearly justified.[2,31]

## Predictive purpose

A structured "roadmap" approach to prediction questions follows the same general steps as outlined above. Full clarity of specification of the question requires definition of the target population, and of the outcome and predictors, with the identification stage, prior to analysis planning, requiring consideration of sampling/recruitment and outcome and predictor measurement quality. Prediction problems invariably involve multiple predictors and seek to develop an algorithm (i.e. in our usage, a procedure) for reliably forecasting the value of $Y$ for individuals for whom only the values of the $X$'s are available. Prediction tasks often (but not always) concern binary outcomes: for example, the question of interest may involve developing an algorithm or formula with which to predict the risk of dying or the risk of relapse (or cure) in a population of patients with a particular disease, given a set of available covariates.

### *Prediction as a natural task for regression modelling*

Multivariable regression models in the form of (3) are naturally suited to prediction tasks, since they are inherently designed to represent or "predict" the expected value of $Y$ as a function of one or more $X$'s. Developing regression methods for prediction thus provides a natural context for further development of the theory of least-squares estimation (for the mean of a continuous outcome) and for extending this to generalised linear models (with maximum likelihood estimation), such as the canonical logistic regression, for binary outcome variables.

In the latter vein, if the aim is to predict a binary outcome from a set of input variables, the target parameter of interest is the probability of the outcome (equal to its expected value), so the essential task of a predictive analysis is to find ways of developing a formula (or algorithm) that expresses this probability as a function of input variables. If we start with the classical linear regression formulation (2), we quickly observe that the unbounded linear expression is not well suited to representing the variation in the range-restricted probability. A mathematically natural alternative is to *transform* the probability to the log-odds or logistic scale (i.e. use model (3) with $g(\pi) = \text{logit}(\pi)$), and investigate if linear combinations of predictors can be identified to accurately represent the variation in this parameter across the multidimensional domain of (vector) $\boldsymbol{X}$ (according to measurements of predictive performance as outlined below). Thus, a purpose-based curriculum for biostatistics might introduce the basic concepts of logistic regression at an early stage, in the treatment of prediction.

***Building prediction models: general concepts***

Although modern best-practice approaches to prediction modelling employ a range of advanced techniques, in a teaching context it is natural to begin with simple ideas. For example, it can be observed that the simple standard regression specification, as in (3), with parallel linear terms and no interactions, should not be expected to perform optimally in many applications. Assuming the same prediction function for males and females (for example) may result in poor predictive performance. In general, complex predictor functions, at least including nonlinearities and interactions, will be needed, and these concepts can be introduced to students within the prediction task framework. This opens the way to examining the challenge of developing appropriate strategies for prediction model building. Traditional courses in regression – and popular software packages – place considerable emphasis on variable selection methods based on p-value criteria, but these have long been known to be flawed in many ways. In particular, for prediction modelling, such approaches ignore the purpose at hand, of optimising the quality of the predictions produced by the model.

Modern approaches to building prediction models in a regression framework include resampling methods such as bootstrap and cross-validation, and shrinkage estimation, such as the "least absolute shrinkage and selection operator" (lasso), while a principled approach to "best subset of predictors" selection has also recently appeared.[3,32,33] Nowadays, discussion of prediction should also include techniques developed in computer science under the general heading of "machine learning", although there is evidence that these approaches may not provide substantial advantages over regression methods in clinical prediction settings[34]. Of course, this conclusion may not apply in the high-dimensional problems that are increasingly common.

***Validating prediction models***

Methodology for assessing the quality of predictions (often referred to as validation) must focus on evaluating the scale and consequences of prediction errors. Such evaluation can only be done in context, where prediction errors will have real-world costs that need to be understood by the analyst. This is in contrast to traditional statistical teaching of regression, which for continuous outcomes has focused on the understanding of residual variance and classical decompositions of sums of squares. It is of mathematical interest to consider the extent to which the total variability in an outcome may be "removed" by using the prediction model, but the classical concepts of "variance explained" do not assist in determining whether a prediction model is useful. This can only be answered in the context of the specific research question.

However, a range of technical concepts and tools are available for model validation in the more common context of predicting binary outcomes.[3] Overall predictive performance can be quantified in terms of "discrimination" and "calibration". *Discrimination* measures how well the predicted probabilities distinguish between those who experience the outcome and those who do not, depending on the choice of threshold value used to convert predicted probabilities into binary (or indeed multi-category) predictions. *Calibration* measures how well the predicted distribution matches the observed outcome distribution. Again, context-specific information such as the relative costs and benefits of false positive and false negative predictions are important in the practical evaluation and implementation of prediction models.

It should be apparent that using the same data for model development and for evaluation of predictive ability will in general lead to over-optimistic performance measures ("overfitting"). A simple approach to managing this problem, while undertaking *internal* validation (which assesses performance within the same dataset that was used to develop the model), is to split the available data into a "training" set, used to develop the prediction model, and a "validation" set, in which the predictive performance is evaluated. More sophisticated methods such as cross-validation use repeated sample-splitting, in which models are developed on randomly sampled subsets of the data and validated on the remaining data, for multiple such splits, with the results combined appropriately. Ideally there should also be a plan for *external* validation, an assessment of how well the prediction model performs in similar but not identical contexts (whether defined temporally, geographically, or by other aspects of the population definition).

***Regression for prediction: summary***

A key feature of the use of standard regression methods for prediction is that the estimated regression coefficients are rarely if ever useful, beyond determining the prediction algorithm. It is sometimes suggested that the "importance" of an individual predictor can be gauged by examining the size of its regression coefficient, but such interpretation again begs the question of "importance for what purpose?". If the purpose is prediction, then it is unclear how the size of the coefficient indicates its importance; for that we would need to examine predictive performance measures with and without that predictor.

In summary, once crucial design aspects have been considered in the first steps of the roadmap (target population and sample, definition and measurement of outcome and predictors), prediction problems become fertile ground for the development of a considerable toolkit of techniques. However, students and practitioners need continual reminding that the statistical aspects of bias (due to inadequate sampling of the target population and overfitting) and variance (largely due to sample size) remain critical in their potential to impact predictive performance.

## Causal purpose

Hernán and others have argued persuasively that a large proportion of epidemiological and clinical research studies have a causal purpose – they seek to answer a causal or "What if…" question.[16,35,36] Such questions are ideally answered by experimentation, in which treatments or exposures are assigned by the researcher, but of course this is often impossible. A substantial body of literature in epidemiology, social science and elsewhere shows that observational studies can also be used to address such questions, subject to inevitable limitations. Crucially, a vast majority of the published literature in these fields uses regression modelling to address such questions.

A comprehensive account of theory and methods for causal inference will not be possible in any single teaching program, especially acknowledging the various strands that have developed and the many aspects that are still debated. However, we suggest that a systematic introductory course could use a structured "roadmap" approach as we have outlined for other question types, which encompasses what we believe are the three essential key steps to causal inference. (Note that our simplified 3-step "roadmap" is different from, but consistent with, what has been termed "The Roadmap for Causal Inference" by van der Laan and colleagues. [37]) Following this outline has formed the basis of successful short courses that we have delivered to health researchers and biostatisticians, and as will be seen, provides a framework within which the potential role of regression methods to address causal questions may be defined.

### *Defining the causal estimand of interest*

One of the key contributions of the causal inference literature has been in formally defining causal estimands, with the canonical example being the average causal effect. This is defined as a contrast (difference or ratio) between average outcomes that *would have been observed* in the target population with and without the treatment or intervention of interest. This estimand can be formalised mathematically through the use of potential outcomes, which refer to outcomes under alternative intervention scenarios.[38] For example, for a binary 0/1 treatment indicator $X$, the average causal effect can be defined as a contrast, often the difference, between $E(Y^{X=1})$ and $E(Y^{X=0})$ where $Y^{X=x}$ denotes the potential outcome when setting $X = x$.

A useful intuitive way of teaching about this estimand is as the effect that would be obtained in an ideal randomised experiment or trial (i.e. a trial with an infinite and representative sample, perfect adherence and follow-up, and no missing data or measurement error) in which individuals are randomised to $X = 0$ or $X = 1$.[39] Crucially, even when experimental control of treatment allocation is not possible, causal questions can still be helpfully framed in terms of the hypothetical trial that would provide an answer if it were possible to conduct. This is the central tenet of the so-called "target trial" approach[15,40], surrounding which there are different perspectives on how "ideal" the specified target trial should or can be. [41] Regardless, specification of the target trial involves defining the population of interest (by way of eligibility criteria), the treatment or interventions to compare, the assignment procedures, the follow-up period, the primary outcome measure, as well as the causal contrast of interest or effect measure (e.g. the difference in means for a continuous outcome, the risk ratio for a binary outcome, etc). We have found the target trial framework to be a useful device for translating formal causal inference concepts into practice and teaching

### *Considering identifiability assumptions*

Only if certain assumptions hold is the causal effect estimable, meaning that the average causal effect can be expressed as a function of the observable data or, within the target trial framework, that the target trial can be closely emulated with the data at hand. These assumptions are known as *causal identifiability* assumptions. With a representative sample, and in the absence of missing data and measurement error, there are three assumptions that suffice. The first is consistency, which in practice, within the target trial framework, requires positing a well-defined intervention in the target trial specification, and then making sure this corresponds to the intervention measure captured in the data and used in the emulation.[39] The second assumption is (structural) positivity, which means that everyone in the population should be able to receive each value of the treatment or exposure. In practice, if using the target trial framework, this requires careful consideration of the eligibility criteria and interventions specified, as well as considering potential random violations of this condition in the actual data used in the emulation (more on sample size issues later). The third assumption is conditional exchangeability given a selected set of confounders $\boldsymbol{Z}$, meaning the treatment or exposure groups are in essence comparable within strata of the confounders. In practice this requires careful consideration of confounding paths to select a sufficient adjustment set.[39,42] If these assumptions hold, then the so-called "g-formula" applies:

$$E(Y^{X=x}) = \sum_{\boldsymbol{z}} E(Y|X = x, \boldsymbol{Z} = \boldsymbol{z})\, P(\boldsymbol{Z} = \boldsymbol{z}); \quad (5)$$

that is, the average causal effect can be expressed as a function of the observable data. Note that in the context of an ideal randomised trial, we have:

$$E(Y^{X=x}) = E(Y|X = x)\,. \quad (6)$$

To select confounders and examine other bias risks (selection and measurement bias), it is helpful to use a causal diagram or directed acyclic graph (DAG) to depict assumed relationships between the analysis variables and other factors that may be relevant, for instance because of their role either as confounders, mediators or so-called colliders.[43-45]. In the presence of missing data, further assumptions are required.[46,47] This process based on the DAG will produce a list of variables that need to be *adjusted for,* in order to identify the causal effect, while also avoiding the risk of bias that can be induced by *over-adjustment*.[48]

***Estimate the causal effect of interest***

Obtaining estimates that are adjusted for confounders and other factors selected is a complex concept, especially in the context of time-varying treatments. Regression provides a specific approach to obtain causal effect estimates with adjustment that is suitable only for the point (i.e. single time) treatment setting and corresponds to *conditioning on* the values of the adjustment variables, while assuming that the causal effect of interest is constant within all subsets of the population defined by combinations of these.

In teaching these ideas, it is helpful to point out that the result in (6) implies that, under the identifiability assumptions, the simplest possible regression model (1), with $X$ a binary indicator for treatment or exposure, provides a parametric representation of the causal effect of interest (as $\beta_1$) in an ideal randomised trial. Inference may be performed for an actual trial, if it can be assumed to be a finite version of the ideal (which is rarely plausible), by fitting this regression in statistical packages, just as we outlined above in the context of a descriptive comparison of two sub-populations.

Moving on to the observational study setting, again focusing on the ideal scenario (apart from randomisation), if the causal effect is constant within substrata of $\boldsymbol{Z}$, and again the identifiability assumptions hold, then following (5) the average causal effect in the mean difference scale is equal to:

$$E(Y^{X=1}) - E(Y^{X=0}) = E(Y|X = 1, \boldsymbol{Z} = \boldsymbol{z}) - E(Y|X = 0, \boldsymbol{Z} = \boldsymbol{z})$$

for any value $\boldsymbol{z}$ taken by $\boldsymbol{Z}$. This means that we can estimate the causal effect by fitting a regression model that includes the confounding adjustment variables $\boldsymbol{Z}$ as well as the treatment indicator.

Specifically, beginning for pedagogical purposes with the case of a continuous outcome $Y$ for which the difference in means is the causal effect measure of interest and there is a single binary confounding variable $Z$, a possible regression model is:

$$E(Y|X,Z) = \beta_0 + \beta_1 X + \gamma Z. \tag{7}$$

Here $\beta_1$ captures the average causal effect of interest. Pursuing the details a little further, if the single confounder $Z$ is continuous, the model above would be imposing the default assumption of a common, linear relationship with the outcome for each value of treatment. The latter parametric assumption may or may not be reasonable from a substantive perspective.

The regression formulation readily extends to a larger adjustment set (multiple confounders) and adapts to handling other effect measures. For multiple confounders, the outcome regression specification (7) extends to:

$$E(Y|X,\boldsymbol{Z}) = \beta_0 + \beta_1 X + \gamma_1 Z_1 + \cdots + \gamma_k Z_k \tag{8}$$

where initially $\boldsymbol{Z}$ might represent a vector of $k$ distinct adjustment variables, but more generally may be developed as a specification involving fewer than $k$ covariates but including various “non-linear” terms such as polynomial functions, to represent curved relationships, and product terms to represent interaction effects. Recall that we would only use this model if we continued to consider that the assumption that the causal effect of interest is constant across all the confounder substrata defined by every combination of values of $Z_1, \ldots, Z_k$ is reasonable.

Emphasising the constant-effect assumption and the parametric assumptions encoded in the specification of the confounder adjustments provides an opportunity to mention that there are more general approaches to causal effect estimation that require fewer assumptions, such as g-computation and inverse-probability weighting, and the more recent doubly robust methods.[39,49] All these methods allow for effect heterogeneity in estimating the average causal effect.

Finally, there is the important issue of outcome regression for estimating other causal effect measures, for outcomes for which (8) does not provide a good model. When the outcome is binary, attention often focuses on ratio effect measures such as the risk ratio or odds ratio, although the “linear” risk difference should often also be of interest. It is instructive to demonstrate how standard approaches to the ratio effects extend in a simple way from (7) by introducing an appropriate link function. For instance, the risk ratio can be studied under similar assumptions as above, by considering the following modification of (7) obtaining by taking logs:

$$\log E(Y|X,Z) = \beta_0 + \beta_1 X + \gamma Z. \tag{9}$$

The log link recovers the causal effect of interest here, the risk ratio, which is represented by $\exp\left(\beta_0 + \beta_1 \times 1 + \gamma Z\right) / \exp\left(\beta_0 + \beta_1 \times 0 + \gamma Z\right) = \mathrm{e}^{\beta_1}$, highlighting the now-familiar assumption of a constant causal effect for every value of $Z$. Expression (9) immediately extends to a multiple-confounder version analogous to (8). For practical implementation, we face the challenge of appropriate estimation methods for such models, but these fall into the broad category of "generalised linear models", for which estimation theory and computational tools abound.

Of note, however, the log-link binary outcome regression has been used much less often in practice than logistic regression, which in the causal context provides an approach to estimating the odds ratio exposure effect (using a similar development to that outlined above for the risk ratio).

Several important areas of discussion on the use of regression in causal inference are opened here. First, why choose one effect measure over another? Students of epidemiology traditionally consider this question in rather more detail than students of biostatistics, but many of the issues involved benefit from a strong grasp of mathematical and statistical issues (e.g. effect heterogeneity[50] and collapsibility across strata[51]). Second, once the preferred target effect measure (estimand) has been decided, and identifiability issues considered, what practical problems may arise in its estimation? A serious but under-recognised threat is that of "sparse data bias", which arises when sample size is small or when there are a large number of adjustment variables relative to sample size.[52] Another estimation problem is seen in the form of numerical difficulties that arise in the fitting of outcome regression models for binary effect measures such as risk ratio and risk difference. Approaches to managing these problems may include penalisation methods for sparse data[53] and modified estimation methods for unstable likelihoods[54], while more satisfying alternatives may lie in the use of the more general approaches to causal inference mentioned above. Specifically, doubly robust methods are likely to be of particular importance for addressing the risk of sparsity bias, which can affect g-computation and inverse-probability weighting as well as outcome regression. Doubly robust methods can incorporate machine learning and thus provide a principled way of dealing with such high-dimensional settings[42,55] and they may also present an important avenue for estimation of stratum-specific causal effects that may approximate the unidentifiable subject-specific causal effects that are often of ultimate clinical interest.[56]

## *Discussion*

There is no doubt that regression analysis dominates the landscape of biostatistics in practice, so the importance of establishing strong foundations for its appropriate application seems undeniable. We have pointed in this article to a widespread lack of clarity of purpose in the application of regression methods. Rarely do researchers or their statistician collaborators explicitly clarify their research question as either descriptive, predictive or causal, and there are many examples of multivariable regression being used with the vague aim of identifying "important risk factors" for an outcome of interest. Even when the purpose is clearly identifiable as one of the three types of question, we observe many examples of misuse of regression, as documented in our brief review. In descriptive questions, a reflexive reliance on regression with insufficient awareness of model assumptions often drives the way in which the data are described, with covariate adjustments that may be unwarranted or unhelpful. In prediction questions, analysts will erroneously ascribe meaning to regression coefficients and related sample-dependent inferences. In causal questions, we observe a lack of clarity of a range of concepts, beginning with the key one of the target estimand, but also including confounding and the meaning of adjustment (often shrouded in a hazy concept of "isolating independent effects"), resulting in issues such as the Table 2 fallacy.

We have argued that this abundance of problems with the use of regression analysis stems from the way in which these methods have been traditionally taught, both to professional statisticians and to non-specialists. Students learn very early that the role of the statistical analyst is to "fit the right model", from which interpretation and conclusions will flow. Much emphasis is placed on "goodness of fit" and on methods for model checking, but this tends to reinforce the "true model myth", the notion that identifying the best model is the primary task. Unless, however, one believes that a "true" regression model underlies and is feasible to specify for every problem, which is implausible in health and medical research, then it is scarcely surprising that major problems of usage and interpretation arise.

The concerns raised here relate to broader themes in the history of applied statistics. George Box's famous aphorism "all models are wrong, but some models are useful" may be traced to his 1976 paper "Science and Statistics"[8], which used the career of R.A. Fisher as an exemplar of how statistics contributes to science by being fully engaged in the substance of scientific enquiry. Fisher's work was based in agricultural science and genetics, where he pursued questions that would now categorise as either descriptive or causal. Within the former category he developed the mathematical theory of regression and analysis of variance, while in the latter he was a pioneer of experimental design. Arguably, however, the brilliance and originality of his mathematical work may have inadvertently fuelled a subsequent trend towards greater emphasis on mathematical theory and packaged techniques within the discipline of statistics. In any case, for a whole range of reasons, the mid-20th century saw a body of techniques become codified as widely useful, but in a general sense that lost connection with the specific types of scientific enquiry for which they were relevant. A widely cited reflection on these issues by Leo Breiman (2001)[57] criticised what he identified as a statistical culture that assumed that "the data are generated by a given stochastic data model". In essence, Breiman focused on prediction problems and identified many of the same mistakes and misconceptions that we have described above in this context.

More recently, Shmueli (2010)[58] recognised the three types of empirical question and sought to distinguish between statistical modelling for "explaining" and for "predicting", although she did not base her discussion of causal modelling on explicit definitions of target estimands. Unfortunately, causal inference has until recently been something of a taboo topic in (bio)statistics.[36] The truism that "correlation does not imply causation" has been long emphasised and has no doubt led statisticians to shy away from the notion that any useful statements about causation can be made from non-experimental data. However, as outlined here, recent work has shown how a formal theory based on "potential" or "counterfactual" outcomes, which can be further distilled into the target trial concept, provides a way around the traditional taboo.[35,59,60] Unfortunately, the teaching and practice of mainstream biostatistical methods have largely failed to keep up with these developments, despite considerable advances in the world of epidemiology, in which biostatisticians have played a major role.[43] A recent commentary by Stijn Vansteelandt, marking the 20th anniversary of Breiman's paper, provided a rare pointer in this direction, noting that "many traditional statistical analyses focus more on the model than on the problem one is trying to solve" and observing “a great danger in drawing false conclusions when viewing the fitted model as a representation of the ground truth (i.e., as a data-generating model).”[61]

As highlighted earlier, when referring to a classic textbook on logistic regression, the problems we have identified can be traced back to the way that applied statisticians are trained, which is heavily influenced by a small number of texts. For example, a relatively recent and widely used book providing a broad introduction to regression methods in biostatistics states its purpose as to describe “a family of statistical techniques that we call *multipredictor* regression modeling”.[62] The book goes on to distinguish between different types of “application”, including prediction and “isolating the effect of a single predictor” (causal), but fails to explain the full implications of these different applications for issues of model specification and interpretation. A less traditional text[63] emphasises the importance of purpose and the tentativeness of model specifications, with in-depth discussion on the interpretability of regression coefficients, but it remains ambiguous as to whether a model specification may precede a purpose.

Against this background, we have proposed a substantial rethinking of the way regression analysis is traditionally conceived in statistics. Essentially, we emphasise that a regression model is a provisional simplification of reality that must be specifically constructed for the purpose at hand. That purpose may be either descriptive, predictive, or causal, and the approach to developing and interpreting regression models that are useful within each of these contexts differs substantially. Thus, the usefulness of a model can only be determined by considering whether it addresses a clearly posed descriptive, predictive or causal question, under well specified and plausible assumptions (most of which must rely on substance-matter knowledge). The proposed framework emphasises the commonality between regression methods used for different data types or (in more helpful terms) target parameters; thus, all of the usual catalogue of linear regression, logistic regression and other forms of generalised linear model should be seen as relatively minor variations of each other, the development and application of each primarily driven by the purpose or question at hand rather than by the scale of the outcome variable.

We note connections between the issues discussed here and other persistent problems in the mainstream use of statistical methods, such as the ubiquitous notion that the essence of statistical analysis is to test (point) hypotheses, resulting in dichotomous declarations of differences found (or not), irrespective of the underlying purpose of the research. Many of the papers examined in our brief review exhibit this issue, often even in the descriptive presentation of study groups (typical “Table 1”). In this light, our suggested purpose-focused approach to the teaching of regression methods really should be extended to the teaching of statistical methods more generally. Introductory courses emphasise the distinction between parameter and estimate but spend far too little time on defining the parameter, before a statistical model (“assume the outcome is normally distributed…”) is introduced, generally out of thin air. Parameters need to be defined from research questions, not by way of conventional statistical models. There is invariably no research question underlying the typical Table 1, so there is no role for statistical inference.

Our work is of course not the first to call for improvements in biostatistical practice; such calls can be traced back at least as far as the first issue of *Statistics in Medicine*,[64] with subsequent commentaries tracking continuing if not growing problems alongside the increasing complexity of data and analytic tools.[65,66] Recent efforts to improve standards have focused on the development of consensus guidelines for the reporting of medical research studies, e.g. CONSORT[67] for randomised trials, TRIPOD[68] for prediction studies, and the currently developing TARGET[69] initiative for observational causal studies. Although standards for reporting do not directly guide the tasks of study planning and analysis, they can create useful "guardrails" for guiding improved practice. Calls for an increased focus on problems in science, especially concerning the so-called "replication crisis", have fed growing interest in open science practices such as open access and study registration[70], which are also important to improve practice. Recently a group of cognitive scientists has also called for broader changes in research and teaching, highlighting in general terms a similar theme to ours with respect to statistical methods: "… in the absence of a clear understanding of the nature of a research question, the nature of a possible answer to that question, and how to apply statistical methods to get from a question to an answer, we have cargo-cult inference. That is, something that has the superficial appearance of scientific inference, but lacks the underpinnings required for it to actually be considered scientific."[71]

The approach that we propose here is radical, in the sense of requiring significant change to course curricula (and textbooks) and to long-entrenched habits among many statisticians. We believe this fundamental change is needed to significantly improve the widespread poor practices that we have described. Additionally, unless those at the coalface of biostatistical teaching and practice join the challenge of reform, we are likely to see an increasing gap between the rapid progress of new ideas and new methods for causal inference and prediction modelling, on the one hand, and the mass production of poorly conceived multivariable regression analyses in medical journals, on the other.

In summary, we believe it is time the (bio)statistics profession paid more serious attention to the ways in which key statistical methods are used and abused in practice. Reform is essential to ensure our continuing relevance as engaged collaborators in the pursuit of scientific knowledge.

## *Acknowledgements*

An early version of this paper was presented as the President's Invited Speaker lecture delivered by the first author at the International Society for Clinical Biostatistics (ISCB43) conference, Newcastle, U.K., 22 August, 2022. The authors would like to thank our local ViCBiostat-CEBU causal inference team for their contributions to the development of the short course "Observational studies: modern concepts and analytic methods" (which we have delivered biannually since 2019), upon which this manuscript draws. The Murdoch Children's Research Institute is supported by the Victorian Government's Operational Infrastructure Support Program. This work was supported by funding from the Australian National Health and Medical Research Council (Investigator Grant 2009572 to MMB).

# On the uses and abuses of regression models: a call for reform of statistical practice and teaching

John B. Carlin & Margarita Moreno-Betancur

Murdoch Children's Research Institute & The University of Melbourne

## *Supplementary Material*

### Appendix 1

Additional details of Example 1 (kidney lengths in children with urinary tract infection, with and without pyelonephritis).

**Figure A1.1**

Scatter plot of kidney length vs age for the two groups (red triangle = infected (pyelonephritis), blue circle = uninfected), with superimposed cubic regression lines (imposing constant group mean difference). This figure reproduces Figure 1 from the main text using colour and a better choice of plotting symbols to provide a clearer visualisation. Dashed vertical lines and grey bars also show the crude mean differences by five age groups (0-6 months, 6-12 months, 12-24 months, 24-36 months, 36-60 months)

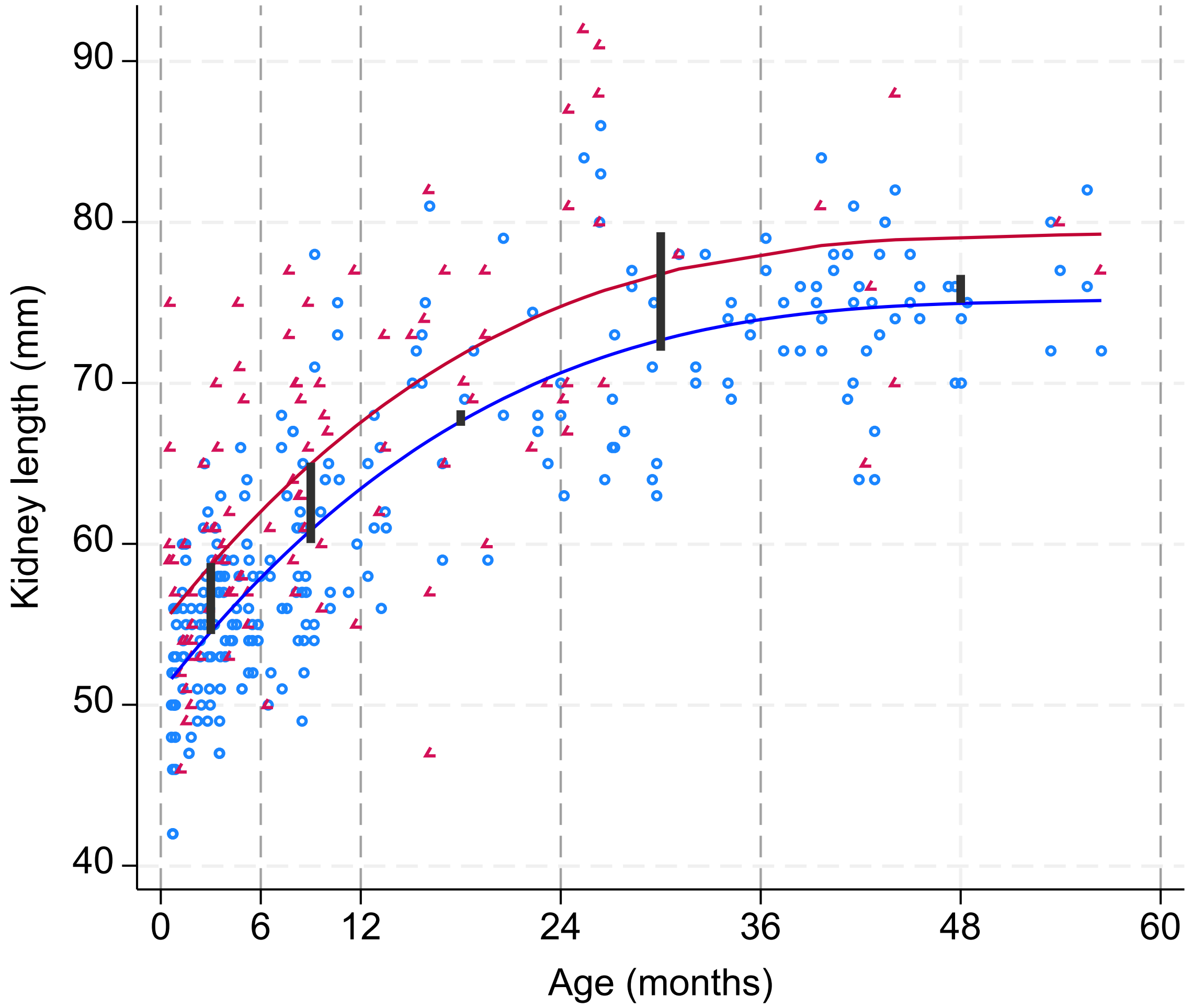

**Table A1.1**

Summary statistics on kidney length by age group, for the infected and uninfected groups.

| | Uninfected | | Infected | | | |
|---|---|---|---|---|---|---|
| Age group | *n* | mean | *n* | mean | difference in means | pooled SD, *s* |
| 0-6 | 104 | 54.4 | 42 | 58.8 | 4.42 | 5.68 |
| 6-12 | 42 | 60.0 | 22 | 65.0 | 5.00 | 7.27 |
| 12-24 | 25 | 67.3 | 17 | 68.3 | 0.96 | 7.28 |
| 24-36 | 32 | 72.0 | 11 | 79.4 | 7.36 | 7.71 |
| 36-60 | 48 | 75.0 | 7 | 76.7 | 1.74 | 4.71 |

**Table A1.2**

Estimates of the difference in mean kidney length between infected and uninfected patients, obtained by methods that make increasingly strong modelling assumptions.

| Method | Estimate of mean difference | Estimated SE |
|---|---|---|
| Simple difference of overall means (equivalent to univariate regression on group indicator) | 2.56 | 1.20 |
| Simple average difference of means across 5 age groups | 3.90 | 0.91 |
| Weighted* average difference of means across 5 age groups | 3.95 | 0.74 |
| Regression-based estimate with adjustment for age groups using 4 indicator variables | 4.14 | 0.73 |
| Regression-based estimate with adjustment for age using cubic polynomial in continuous age | 4.13 | 0.71 |

*weights = inverse variance of age-group-specific mean differences

## Appendix 2

# A brief review of the use of regression methods in contemporary medical research

John B. Carlin & Margarita Moreno-Betancur

Murdoch Children's Research Institute & The University of Melbourne

24-Jan-2024

### Objective

The review aimed, for a series of articles representing "mainstream" clinical research published in three journals in June 2022, to:

1. Determine whether regression analysis was used,
2. Examine whether regression was used for a clear purpose, and if so for what type of purpose (descriptive, predictive, causal),
3. For those papers in which regression analysis was used, assess whether key misuses of regression were apparent.

### Methods

The following three journals were identified from a listing of the top 20 "most influential medical journals" (Jemielniak et al, *J Med Internet Res.* 2019), to represent journals that carry a high proportion of observational clinical studies, relative to the number of randomised trials or laboratory studies: *Pediatrics*, *Neurology* and *BMJ Open*.

We categorised articles published in the (arbitrarily) selected month according to the following questions:

1. Was any form of univariable or multivariable regression analysis reported? (Y/N)
2. If yes to q.1, what was the type of research question?
   (D = descriptive, P = predictive, C = causal, V = vague or imprecisely stated)
3. For articles reporting the results of regression analysis (all Y/N):
   a. If the question was descriptive, was multivariable regression used without a sensible rationale? (For example, for comparisons between subpopulations, was regression used to adjust for covariates without a clear reason? Sensible rationale would mention the use of regression as a tool for smoothing or variance reduction, or standardisation, and not describe regression as controlling or adjusting for a covariate as if it were a confounder.)
   b. If the question was predictive,
      i. Did the development of an appropriate regression model omit consideration of predictive performance (validation) as a criterion for appropriateness of the model?
      ii. Were coefficient estimates interpreted as providing an indication of the "strength of effect" of the corresponding predictors?
   c. If the question was causal,

i. Were there issues in adjustment for confounding: either not done or not using an appropriate method to identify a set of variables that needed to be controlled or adjusted for? (Appropriate methods involve consideration of causal assumptions based on substantive knowledge; inappropriate methods omit any justification for the adjustment set or use statistical tests including "variable selection" in multivariable regression.)
ii. Was there evidence of the "Table 2 fallacy", i.e. interpretation of the coefficients of adjustment covariates as if they represented a causal effect of their own?

d. If the question was vague,
i. Were the authors using multivariable regression to identify "important risk factors" (or similar)?
ii. Was there evidence of the "Table 2 fallacy", i.e. interpretation of the coefficients of covariates in the regression as if they represented causal effects, perhaps in the vein of "interesting" (implicitly causal) associations?

## Results

For each of the three journals, we present a summary table displaying the breakdown of papers reviewed according to the four categories of research purpose, showing within each category the number of articles found to exhibit at least one of the misuses of regression analysis listed in the Methods. The tables give a brief description of the types of misuse identified; further details can be found in notes provided on each article, which follow a full listing in Table A2.5 of all articles and their classification on the review questions (linked by consecutive "study identifiers" assigned to each article within each journal).

**Table A2.1**: Papers using regression analysis (11 of a total of 18 reviewed) in the journal *Pediatrics* (Vol.149, Issue 6, June 2022), classified according to the purpose (type of research question) underlying the analysis.

| Type of research question | Descriptive | Predictive | Causal | Vague/unclear | Total |
|---|---|---|---|---|---|
| ***n*** | 2 | 0 | 6 | 3 | 11 |
| **Problems found** | 1<br>(no justification for adjustment) | | 4<br>(1 univariable, ignoring confounding, 1 unclear about confounding adjustment, 2 problems in method used to identify adjustment set) | 3<br>(all risk factor identification & version of T2 fallacy) | 8 |
| **Study identifiers** (sequential numbering from Table A2.5) | | | | | |
| Misuse: yes | P16 | - | P4,P5,P8,P9 | - | |
| Misuse: no | P3 | - | P1,P6 | P14,P15,P17 | |

**Table A2.2**: Papers using regression analysis (13 of a total of 19 reviewed) in the journal *Neurology* (Vol.98, Issues 23 and 24, June 2022), classified according to the purpose (type of research question: D = descriptive, P = predictive, C = causal, V = vague) underlying the analysis.

| Type of research question | Descriptive | Predictive | Causal | Vague/ unclear | Total |
|---|---|---|---|---|---|
| ***n*** | 1 | 6 | 6 | 0 | 13 |
| **Problems found** | 1 | 3 | 4 | | 8 |
| | (no justification for adjustment) | (3 inappropriate model dev, 1 also interpretation of coeffs) | (3 problems in method used to identify adjustment set, 1 Table 2 fallacy) | | |
| **Study identifiers** (sequential numbering from Table A2.5) | | | | | |
| Misuse: yes | N14 | N1,N12,N5 | N11,N15,N16,N9 | - | |
| Misuse: no | - | N2,N13,N19 | N4,N17 | - | |

**Table A2.3**: Papers using regression analysis (12 of a total of 20 reviewed) in the journal *BMJ Open* (Vol.12, Issue 6, June 2022: the first 20 articles published), classified according to the purpose (type of research question: D = descriptive, P = predictive, C = causal, V = vague) underlying the analysis.

| Type of research question | Descriptive | Predictive | Causal | Vague/unclear | Total |
|---|---|---|---|---|---|
| ***n*** | 2 | 1 | 2 | 7 | 12 |
| **Problems found** | 1 | 1 | 0 | 7 | 9 |
| | (no justification for adjustment) | (interpretation of coeffs) | | (risk factor identification & version of T2 fallacy) | |
| **Study identifiers** (sequential numbering from Table A2.5) | | | | | |
| Misuse: yes | B4 | B8 | - | B1,B3,B5,B12,B15,B17,B18 | |
| Misuse: no | B2 | - | B7,B20 | - | |

## Summary

Overall, we examined 57 papers, in 36 (63%) of which regression methods were used. Among these papers, 25 (69%, or 44% of all papers) exhibited a type of misuse of regression along the lines we have described. Although the mix of types of question was quite different across the three journals (see Table A2.4), the proportion of papers in which misuse could be identified was similar.

**Table A2.4**: Summary of review of clinical research papers focusing on those using regression methods, classified according to the purpose (type of research question: D = descriptive, P = predictive, C = causal, V = vague) underlying the analysis. Values shown are number of articles displaying misuse / total number using regression analysis.

| | *Pediatrics* $n = 18$ | *Neurology* $n = 19$ | *BMJ Open* $n = 20$ | Combined $n = 57$ |
|---|---|---|---|---|
| **Descriptive** | 1/2 | 1/1 | 1/2 | 3/5 |
| **Predictive** | - | 3/6 | 1/1 | 4/7 |
| **Causal** | 4/6 | 4/6 | 0/2 | 8/14 |
| **Vague/unclear** | 3/3 | - | 7/7 | 10/10 |
| **Overall (proportion)** | 8/11 (73%) | 8/13 (62%) | 9/12 (75%) | 25/36 (69%) |
| **Proportion all articles** | 44% | 42% | 45% | 44% |

The major problems identified with the use of regression analysis were:

- 10 papers in which regression was used to answer a poorly specified question, along the lines of "we aimed to identify the important risk factors for *Y*", and results were almost universally presented as a table of estimated coefficients for a multivariable regression model, with these interpreted as providing the "independent effect" of the corresponding factor on the outcome.
- 8 papers that used outcome regression analysis to answer a clearly specified causal question (or questions) but failed to provide a clear rationale for the choice of adjustment variables needed to control for confounding (or failed to acknowledge the risk of confounding bias), and in one case interpreted coefficients of a multivariable model as causal effects.
- 4 papers with the objective of developing a prediction model, among which 3 failed to use an appropriate method for developing their model (i.e. based this on statistical significance of terms in the regression model rather than impact on predictive validity) and 2 interpreted coefficients in a pseudo-causal fashion.
- 3 papers using regression for a descriptive research purpose but including adjustment terms without clear justification.

**Table A2.5: Full listing of all articles reviewed, with coding according to review questions (see Methods)**

Coding: Y=yes, N=no, D = descriptive, P = predictive, C = causal, V = vague or imprecisely stated.

| *ID no.* | *Article* | *Q1* | *Q2* | *Q3a* | *Q3b(i)* | *Q3b(ii)* | *Q3c(i)* | *Q3c(ii)* | *Q3d(i)* | *Q3d(ii)* |
|---|---|---|---|---|---|---|---|---|---|---|
| | ***Pediatrics*** | | | | | | | | | |
| P1. | Whitaker RC, et al. Family Connection and Flourishing Among Adolescents in 26 Countries. *Pediatrics.* 2022;149(6). | Y | C | - | - | - | N | N | - | - |
| P2. | Sharko M, et al. State-by-State Variability in Adolescent Privacy Laws. *Pediatrics.* 2022;149(6). | N | D | | | | | | | |
| P3. | Rahman R, et al. Intimate Partner Violence and the COVID-19 Pandemic. *Pediatrics.* 2022;149(6). | Y | D | N | | | | | | |
| P4. | Rao S, et al. Asthma and the Risk of SARS-CoV-2 Infection Among Children and Adolescents. *Pediatrics.* 2022;149(6). | Y | C | - | - | - | Y | N | - | - |
| P5. | Boutzoukas AE, et al. School Masking Policies and Secondary SARS-CoV-2 Transmission. *Pediatrics.* 2022;149(6). | Y | C | - | | | Y | N | | |
| P6. | Elliott LJ, et al. Vegetarian Diet, Growth, and Nutrition in Early Childhood: A Longitudinal Cohort Study. *Pediatrics.* 2022;149(6). | Y | C | - | - | - | N | N | - | - |
| P7. | Morin L, et al. The Current and Future State of Pediatric Sepsis Definitions: An International Survey. *Pediatrics.* 2022;149(6). | N | D | | | | | | | |
| P8. | Oghalai JS, et al. Cochlear Implants for Deaf Children With Early Developmental Impairment. *Pediatrics.* 2022;149(6). | Y | C | - | - | - | Y | N | - | - |
| P9. | Boyle MH, et al. Physical Activity Opportunities in US Early Child Care Programs. *Pediatrics.* 2022;149(6). | Y | D/C | - | - | - | Y | N | - | - |
| P10. | Khan A, et al. Family Safety Reporting in Medically Complex Children: Parent, Staff, and Leader Perspectives. *Pediatrics.* 2022;149(6). | N | D | | | | | | | |
| P11. | Guez-Barber D, et al. Differentiating Bell's Palsy From Lyme-Related Facial Palsy. *Pediatrics.* 2022;149(6). | N | D/P | | | | | | | |
| P12. | Liao F-M, et al. Direct Bilirubin and Risk of Biliary Atresia. *Pediatrics.* 2022;149(6). | N | D | | | | | | | |
| P13. | Gupta K, et al. Dextrose Gel for Neonates at Risk With Asymptomatic Hypoglycemia: A Randomized Clinical Trial. *Pediatrics.* 2022;149(6). | N | C | | | | | | | |
| P14. | Boghossian NS, et al. Transfer Patterns of Very Low Birth Weight Infants for Convalescent Care. *Pediatrics.* 2022;149(6). | Y | V | - | - | - | - | - | Y | Y |
| P15. | Aubert AM, et al. Movement Difficulties at Age Five Among Extremely Preterm Infants. *Pediatrics.* 2022;149(6). | Y | D/V | - | - | - | - | - | Y | Y |
| P16. | de Almeida MFB, et al. Translating Neonatal Resuscitation Guidelines Into Practice in Brazil. *Pediatrics.* 2022;149(6). | Y | D | Y | - | - | - | - | - | - |
| P17. | Price JJ, et al. Cardiovascular Risk Factors and Target Organ Damage in Adolescents: The SHIP AHOY Study. *Pediatrics.* 2022;149(6). | Y | V | - | - | - | - | - | Y | Y |
| P18. | Prins S, et al. How Physicians Discuss Uncertainty With Parents in Intensive Care Units. *Pediatrics.* 2022;149(6). | N | D | | | | | | | |
| | ***Neurology*** | | | | | | | | | |
| N1. | Faizy TD, et al. The Cerebral Collateral Cascade. *Comprehensive Blood Flow in Ischemic Stroke.* 2022;98(23):e2296-e2306. | Y | P | - | Y | Y | - | - | - | - |
| N2. | Yang S, et al. Development of a Model to Predict 10-Year Risk of Ischemic and Hemorrhagic Stroke and Ischemic Heart Disease Using the China Kadoorie Biobank. *Neurology.* 2022;98(23):e2307-e2317. | Y | P | - | N | N | - | - | - | - |
| N3. | Stefanetti RJ, et al. L-Arginine in Mitochondrial Encephalopathy, Lactic Acidosis, and Stroke-like Episodes. *A Systematic Review.* 2022;98(23):e2318-e2328. | N | | | | | | | | |
| N4. | Wiggs KK, et al. Maternal Serotonergic Antidepressant Use in Pregnancy and Risk of Seizures in Children. *Neurology.* 2022;98(23):e2329-e2336. | Y | C | - | - | - | N | N | - | - |
| N5. | Gross WL, et al. Prediction of Naming Outcome With fMRI Language Lateralization in Left Temporal Epilepsy Surgery. *Neurology.* 2022;98(23):e2337-e2346. | Y | P | - | Y | N | - | - | - | - |
| N6. | Morris C, et al. Outcomes That Matter to Adolescents With Continuous Headache Due to Chronic Migraine and Their Parents. *A Pilot Survey Study.* 2022;98(23):e2347-e2355. | N | | | | | | | | |
| N7. | Thomas FP, et al. Randomized Phase 2 Study of ACE-083 in Patients With Charcot-Marie-Tooth Disease. *Neurology.* 2022;98(23):e2356-e2367. | N | | | | | | | | |
| N8. | Molimard A, et al. Rituximab Therapy in the Treatment of Juvenile Myasthenia Gravis. *The French Experience.* 2022;98(23):e2368-e2376. | N | | | | | | | | |
| N9. | Raj R, et al. Risk of Dementia After Hospitalization Due to Traumatic Brain Injury. *A Longitudinal Population-Based Study.* 2022;98(23):e2377-e2386. | Y | C | - | - | - | N | Y | - | - |
| N10. | Gool JK, et al. Data-Driven Phenotyping of Central Disorders of Hypersomnolence With Unsupervised Clustering. *Neurology.* 2022;98(23):e2387-e2400. | N | | | | | | | | |
| N11. | Vitkova M, et al. Association of Latitude and Exposure to Ultraviolet B Radiation With Severity of Multiple Sclerosis. *An International Registry Study.* 2022;98(24):e2401-e2412. | Y | D/C | Y | | | Y | N | - | - |
| N12. | Yoon EJ, et al. Brain Metabolism Related to Mild Cognitive Impairment and Phenoconversion in Patients With Isolated REM Sleep Behavior Disorder. *Neurology.* 2022;98(24):e2413-e2424. | Y | D/P | Y | Y | Y | - | - | - | - |
| N13. | Petersen KK, et al. Predicting Amyloid Positivity in Cognitively Unimpaired Older Adults. *A Machine Learning Approach Using A4 Data.* 2022;98(24):e2425-e2435. | Y | P | - | N | N | - | - | - | - |
| N14. | Whitwell JL, et al. Investigating Heterogeneity and Neuroanatomic Correlates of Longitudinal Clinical Decline in Atypical Alzheimer Disease. *Neurology.* 2022;98(24):e2436-e2445. | Y | D | Y | - | - | - | - | - | - |
| N15. | Sible IJ, et al. Visit-to-Visit Blood Pressure Variability and CSF Alzheimer Disease Biomarkers in Cognitively Unimpaired and Mildly Impaired Older Adults. *Neurology.* 2022;98(24):e2446-e2453. | Y | C | - | - | - | Y | N | - | - |
| N16. | Tarko L, et al. Racial and Ethnic Differences in Short- and Long-term Mortality by Stroke Type. *Neurology.* 2022;98(24):e2465-e2473. | Y | C | - | - | - | Y | N | - | - |
| N17. | Cai M, et al. Association of Ambient Particulate Matter Pollution of Different Sizes With In-Hospital Case Fatality Among Stroke Patients in China. *Neurology.* 2022;98(24):e2474-e2486. | Y | C | - | - | - | N | N | - | - |
| N18. | Grindegård L, et al. Association Between EEG Patterns and Serum Neurofilament Light After Cardiac Arrest. *A Post Hoc Analysis of the TTM Trial.* 2022;98(24):e2487-e2498. | N | | | | | | | | |
| N19. | Abdallah C, et al. Clinical Yield of Electromagnetic Source Imaging and Hemodynamic Responses in Epilepsy. *Validation With Intracerebral Data.* 2022;98(24):e2499-e2511. | Y | P | - | N | N | - | - | - | - |
| | ***BMJ Open*** | | | | | | | | | |
| B1. | Cullen P, et al. Returning to the emergency department: a retrospective analysis of mental health re-presentations among young people in New South Wales, Australia. *BMJ Open.* 2022;12(6):e057388. | Y | P/V | - | - | - | - | - | Y | Y |
| B2. | Gardner LA, et al. Lifestyle risk behaviours among adolescents: a two-year longitudinal study of the impact of the COVID-19 pandemic. *BMJ Open.* 2022;12(6):e060309. | Y | D | N | - | - | - | - | - | - |
| B3. | Bradfield OM, et al. Vocational and psychosocial predictors of medical negligence claims among Australian doctors: a prospective cohort analysis of the MABEL survey. *BMJ Open.* 2022;12(6):e055432. | Y | V | - | - | - | - | - | Y | Y |
| B4. | Rogers A, et al. Adverse events and overall health and well-being after COVID-19 vaccination: interim results from the VAC4COVID cohort safety study. *BMJ Open.* 2022;12(6):e060583. | Y | D | Y | - | - | - | - | - | - |
| B5. | Ha TN, et al. Trend in CT utilisation and its impact on length of stay, readmission and hospital mortality in Western Australia tertiary hospitals: an analysis of linked administrative data 2003–2015. *BMJ Open.* 2022;12(6):e059242. | Y | V | - | - | - | - | - | Y | Y |
| B6. | Alharbi K, et al. Understanding the implementation of interventions to improve the management of frailty in primary care: a rapid realist review. *BMJ Open.* 2022;12(6):e054780. | N | | | | | | | | |
| B7. | Vijayaraghavan BKT, et al. Hydroxychloroquine plus personal protective equipment versus personal protective equipment alone for the prevention of laboratory-confirmed COVID-19 infections among healthcare workers: a multicentre, parallel-group randomised controlled trial from India. *BMJ Open.* 2022;12(6):e059540. | Y | C | - | - | - | - | - | N | N |
| B8. | Ma Y, et al. Development of a novel predictive model for a successful stone removal after flexible ureteroscopic lithotripsy based on ipsilateral renal function: a single-centre, retrospective cohort study in China. *BMJ Open.* 2022;12(6):e059319. | Y | P | - | N | Y | - | - | - | - |
| B9. | Sania A, et al. Rapid antigen testing by community health workers for detection of SARS-CoV-2 in Dhaka, Bangladesh: a cross-sectional study. *BMJ Open.* 2022;12(6):e060832. | N | | | | | | | | |
| B10. | Wilander H, et al. Nationwide observational study of incidence, management and outcome of spontaneous coronary artery dissection: a report from the Swedish Coronary Angiography and Angioplasty register. *BMJ Open.* 2022;12(6):e060949. | N | | | | | | | | |
| B11. | Puértolas OC, et al. Evaluating the controlled reopening of nightlife during the COVID-19 pandemic: a matched cohort study in Sitges, Spain, in May 2021 (Reobrim Sitges). *BMJ Open.* 2022;12(6):e058595. | N | | | | | | | | |
| B12. | Tiwari L, et al. Clinicodemographic profile and predictors of poor outcome in hospitalised COVID-19 patients: a single-centre, retrospective cohort study from India. *BMJ Open.* 2022;12(6):e056464. | Y | V | - | - | - | - | - | Y | Y |
| B13. | Bywall KS, et al. Making space for patients’ preferences in precision medicine: a qualitative study exploring perspectives of patients with rheumatoid arthritis. *BMJ Open.* 2022;12(6):e058303. | N | | | | | | | | |
| B14. | Lurie P, et al. COVID-19 vaccine misinformation in English-language news media: retrospective cohort study. *BMJ Open.* 2022;12(6):e058956. | N | | | | | | | | |
| B15. | Muhammad T, et al. Socioeconomic and health-related inequalities in major depressive symptoms among older adults: a Wagstaff's decomposition analysis of data from the LASI baseline survey, 2017–2018. *BMJ Open.* 2022;12(6):e054730. | Y | V | - | - | - | - | - | Y | Y |
| B16. | Moussallem M, et al. Evaluating the governance and preparedness of the Lebanese health system for the COVID-19 pandemic: a qualitative study. *BMJ Open.* 2022;12(6):e058622. | N | | | | | | | | |
| B17. | Silwal PR, et al. Understanding geographical variations in health system performance: a population-based study on preventable childhood hospitalisations. *BMJ Open.* 2022;12(6):e052209. | Y | V | - | - | - | - | - | Y | Y |
| B18. | Silver NA, et al. Charming e-cigarette users with distorted science: a survey examining social media platform use, nicotine-related misinformation and attitudes towards the tobacco industry. *BMJ Open.* 2022;12(6):e057027. | Y | V | - | - | - | - | - | Y | Y |
| B19. | Yu H, et al. RECIST 1.1 versus mRECIST for assessment of tumour response to molecular targeted therapies and disease outcomes in patients with hepatocellular carcinoma: a systematic review and meta-analysis. *BMJ Open.* 2022;12(6):e052294. | N | | | | | | | | |
| B20. | Adhikari I, et al. Association of *Chlamydia trachomatis* infection with cervical atypia in adolescent women with short-term or long-term use of oral contraceptives: a longitudinal study in HPV vaccinated women. *BMJ Open.* 2022;12(6):e056824. | Y | C | - | - | - | N | N | - | - |

## Notes on articles reviewed, to support coding according to review questions

P3. Paper documents an increase in intimate partner violence referrals after the start of the COVID-19 pandemic. Description of "before" and "after" with a univariable regression used to assess the size of change between periods.

P4. Examines whether asthma is associated with increased risk of COVID-19. Uses propensity score matching: "Because we identified substantial differences in patient characteristics and SARS-CoV-2 testing rates by asthma status, we used propensity score matching to construct a cohort of children with asthma and control children without asthma closely matched on these variables." Don't mention confounding as such and not clear about the need for the variables used to account for all confounding between asthma status and risk of COVID. (Y,N)

P5. District-level analysis of incidence of SARS-CoV-2 infections, comparing 3 masking policy regimes. No mention of potential confounding. Regression seems to have been used only for unadjusted comparisons. (Y,N)

P6. Focus on causal effect of "vegetarian diet" with multiple adjustments "determined a priori from literature review".

P8. Aim was to evaluate the effect of cochlear implants by comparing 3 cohorts who had differential access to this treatment. No adjustments were made for potential confounding, on the basis that they "found no differences among the cohorts on the basis of study site, sex, household income, race, or mother's education level (Table 1)." Regression used to estimate "trajectories" in outcomes over time allowing for interaction of cohort with age. (Y,N)

P9. First part of paper is descriptive analysis of physical activity in early childhood education centres. Final part investigates the (causal) effects of potential "barriers to physical activity" using regression to adjust for covariates. No explicit reason given for chosen list of adjustment variables nor consideration of potential unmeasured confounding. (Y,N)

P13. RCT with no adjustment.

P14. Aims and design were unclear. A large part of the analysis focuses on descriptive presentation, but they use multiple regression to identify "risk factors" with Figure 2 apparently a version of the Table 2 fallacy. Many "significant" risk factors but no mention of possibility of interactions.

P15. Initial aim was to describe prevalence of different levels of disability in a cohort of children with cerebral palsy. Second part "aimed […] to identify sociodemographic, perinatal, and neonatal risk factors associated with movement difficulties." (Intro) "We produced 3 models to measure the association of sociodemographic, perinatal, and neonatal variables with the probability of being at risk or having significant movement difficulties using multinomial logistic regression…" (Methods) Table 3 exhibits Table 2 fallacy.

P16. Aim was to describe the uptake over time (2014-2020) of best-practice guidelines for neonatal care of preterm newborns (23-31 weeks gestation). Slightly unclear but logistic regression used to estimate change (as OR) per year in use of specific procedures with adjustment for "center and year" (confusing because year seemed to be the factor of interest in the analysis).

P17. "To address the primary research question, generalized linear models were constructed to determine if CVRFs in combination with other known risk factors of end organ changes, including demographic, biometric and laboratory analyses, could predict TOD in our population." Despite this wording, the primary interest appeared to be causal (last sentence of Conclusion: "Future studies should address whether amelioration of these CVRFs in the young averts development of TOD and adult cardiovascular disease.") N.B. Backward elimination was used to determine adjustment set.

N1. Cohort study looking at prediction or risk stratification according to "CCC" status. "Primary outcome analysis" m-v binary logistic regression model with CCC status (3 levels) adjusted for covariates selected by backwards elimination. Adjusted ORs presented as indicating strength of prediction.

N4. Excellent example of causal inference using outcome regression.

N5. Aimed to determine whether fMRI measure predicted outcome. Main result was that "a hierarchical multivariable regression model showed that fMRI added significant independent predictive value beyond the other predictors" (not clear what "hierarchical" meant here), based on first reducing covariates according to "significance" and then comparing R-sq before and after adding fMRI. Small sample size n=81.

N9. Primary analysis focused on causal effect of TBI on risk of dementia with appropriate adjustment for confounders, but results presented in Table 2 and Figure 3 exhibit "Table 2 fallacy".

N11. Aimed to determine the "effect" of latitude on multiple sclerosis severity. Main analysis was regression adjusted for "confounders" but no discussion of what this meant or why this adjustment set was chosen. If aim descriptive then adjustment unwarranted, but if aim causal (latitude as a proxy for some possible intervention?) then choice of adjustments seemed odd because most of them followed the "exposure" to latitude.

N12. Aimed to stratify patients with rare disorder according to risk of progression to Parkinson's or dementia. Used Cox regression for four putative predictors, 3 selected by screening of image-based measures of brain glucose metabolism, adjusted for age and disease duration. Meaning of adjusted HRs?

N13. Prediction modelling using machine learning.

N14. Describing patterns of clinical assessment measures for patients with two neurological diseases, using "linear mixed-effects model" apparently to estimate mean at baseline and mean rate of change. Adjustment for age at baseline not explained.

N15. "Bayesian linear growth modeling with the brms package […] in R investigated the role of BPV, APOE $\varepsilon 4$, and the passage of time on CSF AD biomarker levels." Various interactions and adjustments with unclear rationale.

N16. Effects of race on mortality – arguably a causal question, with authors interested in finding explanations for the disparities, but choice of adjustment variables unclear and included "post-exposure" variables such as comorbidities.

N17. Continuous exposure, attempted confounder adjustment.

N19. Development of MRI measures to predict postsurgical outcome in epilepsy (small sample, informal prediction). Simple univariable regression used in describing results.

B1. Aims were to identify “key characteristics associated with higher risk of […] re-presentation.” Used both univariate and m-v regression, concluding that the latter “resulted in the same set of significantly associated characteristics and did not change the observations and interpretations made with the univariate regression analyses” (?)

B2. Comparing health-related behaviours in a cohort before and during the pandemic. Used GLMMs to estimate prevalence ratios (random effects for student and school) with no covariate adjustment. Some acknowledgement of potential confounding if looking for causal interpretation.

B3. Multivariate logistic regression for risk of “medical negligence”. “Our multivariate analysis identified a number of significant demographic risk factors associated with being sued”. Discussion exhibits Table 2 fallacy.

B4. Descriptive analysis with K-M estimation of proportions over time, but Cox regression used to compare between vaccines adjusted for age, with no explanation for adjustment.

B5. Two major analyses: (1) multivariable Poisson regression models used “to identify factors associated with each classification of CT use” – see Table 2 (fallacy); (2) compared outcomes between CT exposure groups over time using some form of g-computation (predictive margins)

B8. Made some use of cross-validation in setting tuning parameter for LASSO, and mentioned AUC and calibration, but Table 2 and related text presented coefficient estimates as indicating strength of “independent prediction”

B12. “Binary logistic regression was done to identify predictors of mortality, and multivariable regression analysis was done to control potential confounders.” Table 4 exhibits Table 2 fallacy.

B15. “In this study, […] objective of finding the association between socioeconomic and health status and depression among older adults and explore the contributing factors in the inequalities in late-life depression”. M-v logistic regression presented in Table 3 and interpreted as showing factors “significantly associated with” outcome. Also combined regression coefficients to produce “Wagstaff’s Concentration Index” (?).

B17. Multilevel logistic regressions used – combination of trying to quantify geographic variability (adjusted…) along with determining “effects” of “key sociodemographic, economic, geographical and health system characteristics”. See Table 4 and text referring to for version of Table 2 fallacy.

B18. Using m-v regression models to investigate “the extent to which use of specific social media platforms are associated with [outcomes]”. Tables 2 and 3 appear to present “mutually adjusted” coefficient estimates, interpreted in the text as causal effects.